\documentclass{elsart}

\usepackage{graphicx}
\usepackage{epsfig,rotating}
\usepackage[dvips]{color}

\def\be{\begin{equation}}
\def\ee{\end{equation}}
\def\ba{\begin{eqnarray}}
\def\ea{\end{eqnarray}}

\def\lsim{\raise0.3ex\hbox{$\;<$\kern-0.75em\raise-1.1ex\hbox{$\sim\;$}}}
\def\gsim{\raise0.3ex\hbox{$\;>$\kern-0.75em\raise-1.1ex\hbox{$\sim\;$}}}
\def\for{\qquad {\rm for} \qquad}
\def\theta{\vartheta}

\begin{document}

\begin{frontmatter}
\title{Clustering  of ultra-high energy cosmic ray arrival directions
  on medium scales}

\author[NTNU]{M.~Kachelrie\ss}
\author[APC,INR]{and D.V.~Semikoz}

\address[NTNU]{Institutt for fysikk, NTNU Trondheim, N--7491 Trondheim,
  Norway} 
\address[APC]{APC, Coll\`ege de France, 
11, pl. Marcelin Berthelot, Paris 75005, France}
\address[INR]{INR RAS, 60th October Anniversary prospect 7a,
  117312 Moscow, Russia}

\date{20 December 2005}

\begin{abstract}
The two-point autocorrelation function  of ultra-high energy cosmic
ray (UHECR) arrival directions has a broad maximum around 25 degrees, 
combining the data with energies above $4\times 10^{19}$~eV (in the HiRes 
energy scale) of the  HiRes stereo, AGASA, Yakutsk and SUGAR experiments.
This signal is not or only marginally present analyzing events of a single 
experiment, but becomes  significant when data from several experiments are 
added. Both the energy dependence of the signal and its angular
scale might be
interpreted as first signatures of the large-scale structure of UHECR
sources and of intervening magnetic fields.

\begin{small}
\end{small}
\end{abstract}
\end{frontmatter}

\section{Introduction}
The sources of ultra-high energy cosmic rays (UHECR) are despite of
more than 40~years of research still unknown. Main obstacle for doing
charged particle astronomy are deflections of the primaries in the
Galactic and extragalactic magnetic fields. While the magnitude and
the structure of extragalactic magnetic fields are to a large extent
unknown, already 
deflections in the Galactic magnetic field alone are large enough to
prevent UHECR astronomy if the primaries are heavy nuclei~\cite{nuc,KST05}. 
Assuming
optimistically that the primaries are protons, typical deflections in
the Galactic magnetic field are around five degrees in most part of
the sky at $E=4\times 10^{19}$~eV~\cite{KST05}. Therefore, it might be
possible to perform charged particle astronomy, if moreover deflections
in extragalactic magnetic fields are sufficiently small.

This scenario can be divided in two quite different sub-cases: 
In the first one, a small number of point sources results in
small-scale clusters of arrival directions around or near the true
source positions. Accumulating enough events, the identification of
sources will become possible using e.g.\ correlation studies. Various
studies have been pursued in this direction~\cite{ssc,corr}. In the
second sub-case, the number of point sources is too large to receive
two or more UHECRs from the same source with the present
statistics. However, the sky density of UHECRs reflects the large-scale 
structure of the sources and, possibly, of the intervening 
magnetic fields~\cite{lss,napoli}.  
Therefore, the measured UHECR distribution is anisotropic 
and over-/underdense regions exist that reflect the angular size of up-to
15--20 degrees of  typical structures in the galaxy distribution.
Obviously, Nature might have chosen a mixture of
these two extreme possibilities: The vast majority of UHECR sources
might produce only singlet events, while a subclass of sources with
extreme luminosity might be detectable as point sources via
small-scale clustering studies. Furthermore, point sources might be
easier to identify at the highest energies, if the number density of
sources decreases with  the maximal energy $E_{\rm max}$ to which they
can accelerate as argued in Ref.~\cite{spec}.

In this work, we study the arrival direction distributions of the
UHECRs, putting emphasis in contrast to most earlier studies on
intermediate angular scales. 
Since these two-dimensional  distributions average three-dimensional 
structures (with typical scale $L$) over the mean free path $l$ of UHECRs, 
no anisotropies reflecting the large-scale structure of sources are expected 
for $l\gg L$. To obtain an optimal compromise between the number of
events used, the mean free path $l$ of UHECRs and deflections in magnetic 
fields, it is important to use 
a consistent energy scale when combining different experiments.
Therefore, we discuss first in Sec.~\ref{data} the publicly available
UHECR data and how we rescale the data of different experiments to a
common energy scale. In Sec.~\ref{analysis}, we analyze then the
arrival direction distributions of the used UHECRs and discuss in Sec.~4
our results, before we summarize in Sec.~\ref{sum}.

\section{UHECR data sets and their energy scale}
\label{data}

The available data set of UHECR events consists of
\begin{enumerate}
\item
The publicly available AGASA data set until May 2000 from
Ref.~\cite{AG}, consisting of 57~events with $E\geq 4\times 10^{19}$~eV
and zenith angle $\theta\leq 45^\circ$. The exposure of the AGASA experiment 
is $4.0\times 10^{16}\:$m$^2\:$s sr~\cite{AG}.
\item 
The Yakutsk data as presented at the ICRC 2005~\cite{YK}: 34 events
with energy $E\geq 4\times 10^{19}$~eV and zenith angle $\theta\leq
60^\circ$. The corresponding exposure is
$1.8\times 10^{16}\:$m$^2\:$s sr~\cite{YK}.
\item
The SUGAR data with energy above $E\geq 1\times 10^{19}$~eV 
and zenith angle $\theta\leq 70^\circ$ from Ref.~\cite{SG}. We follow 
Ref.~\cite{full} and use  $4.0\times 10^{16}\:$m$^2\:$s sr as 
approximate exposure for $\theta=55^\circ$.
From this value, we rescale the exposure  approximately  to 
$3.0\times 10^{16}\:$m$^2\:$s sr for $\theta\leq 45^\circ$ and to
$5.3\times 10^{16}\:$m$^2\:$s sr for $\theta\leq 70^\circ$, respectively. 
\item 
The HiRes stereo data set: We deduce the arrival directions of the
events with energy $E\geq 1\times 10^{19}$~eV from
Ref.~\cite{Abbasi:2004ib}.  Then we divide the data set into events
with energy in the range $E=(1-2)\times 10^{19}$,  $E=(2-4)\times 10^{19}$
and $E\geq 4\times 10^{19}$~eV using Ref.~\cite{cris}. 
Its total exposure is $4.6\times 10^{16}\:$m$^2\:$s sr~\cite{Abbasi:2004ib}.
\item
From the Volcano Ranch, Haverah Park, Flye's Eye experiments no detailed
informations are available  about their events. Therefore, we can use
only the events with $E>10^{20}$~eV 
for which the arrival directions are given in Ref.~\cite{NW00}: four
events from Haverah Park, and one both from Volcano Ranch and Flye's Eye. 
\end{enumerate}

We use as angular acceptance $\eta(\delta)$ of a ground array experiment
at geographical latitude $b$ that observes showers with maximal zenith
angle $\theta_{\max}$ (cf., e.g., Ref.~\cite{som}) 
\be  \label{cosz}
 \eta(\delta) \propto \int_0^{\alpha_{\rm \max}} d\alpha\cos(\theta)
              \propto \left[ \cos(b)\cos(\delta)\sin(\alpha_{\rm \max}) +
                      \alpha_{\rm \max}\sin(b)\sin(\delta) \right] \,,
\ee
where
\be
 \alpha_{\rm \max} = \left\{ \begin{array}{l}
             \arccos(\xi)  \for -1\leq\xi\leq 1 \,,\\
             \pi           \hspace{1.4cm}\for \xi< -1 \,,\\
              0            \hspace{1.4cm} \for \xi> 1 \,,
             \end{array} \right.
\ee
and
\be
 \xi = \frac{\cos(\theta_{\max})-\sin(b) \sin(\delta)}
            {\cos(b) \cos(\delta)} \,.
\ee
In contrast to ground arrays, the exposure of Hires in right ascension
R.A. is non-uniform. Therefore, we sample the exposure of HiRes from Fig.~2
of Ref.~\cite{Abbasi:2004ib}.

The absolute energy scale of each experiment has a rather large
uncertainty. To reproduce correctly spectral features like the dip,
the energies $E$ given by the experiments have to be shifted to new
energies $E'$~\cite{dip}. First, we assume that the normalization of the HiRes
stereo spectrum  is consistent with the one of HiRes in monocular mode, 
following Ref.~\cite{c2cr}.
In Ref.~\cite{KS03}, we compared the SUGAR data given in
Ref.~\cite{SG} to the energy spectrum measured by the AGASA and HiRes
experiments. We found that rescaling the SUGAR energies calculated
with the Hillas prescription by 15\% downwards, $E'=E_{\rm
  Hillas}/1.15$, makes their data consistent with the ones from
AGASA. The same is true for the HiRes spectrum if the energy is 
rescaled up-wards by 30\%. In contrast to Ref.~\cite{KS03}, we fix the
energy scale by the HiRes mono data in the present work.  Therefore,
we have to shift the AGASA data by 30\% downwards, and the SUGAR data
by 50\% downwards.  According to Ref.~\cite{YK}, the Yakutsk energy
scale is systematically 15-20\% above the AGASA  energy scale. Thus,
in order to match the Yakutsk data to the HiRes energy scale we 
rescale all energies of UHECR events of Ref.~\cite{YK} by 50\% downwards.

\begin{figure}
\epsfig{file=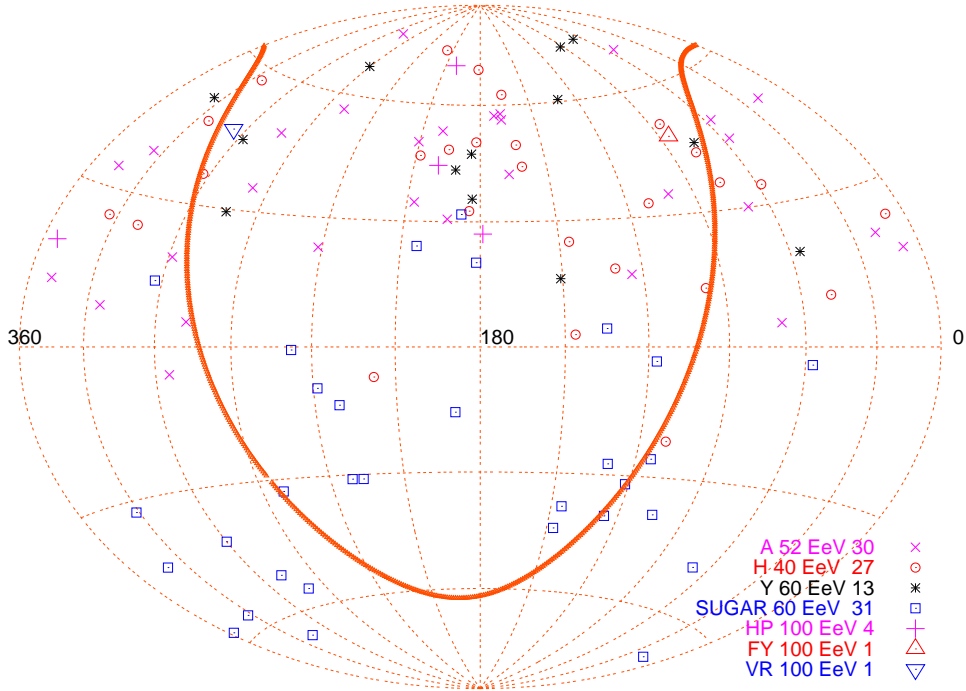,width=0.7\textwidth,angle=270}
\caption{
Skymap of the UHECR arrival directions of events with rescaled energy
$E'>4\times 10^{19}$~eV in equatorial coordinates;  
magenta crosses--30 Agasa (A) events with $E>5.2\times 10^{19}$~eV, 
red circles--27 HiRes (H) events with $E>4\times 10^{19}$~eV, 
black stars--13 Yakutsk (Y) events with $E>6\times 10^{19}$~eV,
blue boxes--31 Sugar (S) events with $E>6\times 10^{19}$~eV, 
magenta crosses--4 Haverah Park (HP) events with $E>10^{20}$~eV, 
red triangle--one Flye's Eye (FY) event with $E>10^{20}$~eV, 
blue triangle--Volcano Ranch (VR) event  with $E>10^{20}$~eV.
\label{skymap1}}
\end{figure}

\begin{figure}
\epsfig{file=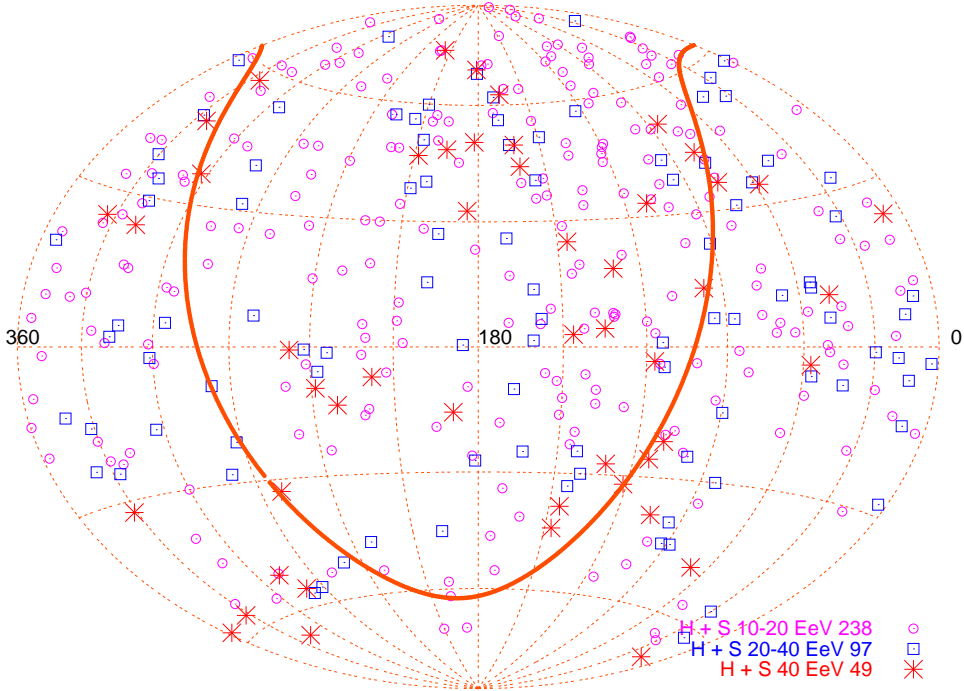,width=0.7\textwidth,angle=270}
\caption{
Skymap of the UHECR arrival directions in equatorial coordinates 
from Hires and SUGAR in three different energy bins, $E=(1-2)\times
10^{19}$ (magenta, small circles),  $E=(2-4)\times 10^{19}$ (blue, medium
box) and $E\geq 4\times 10^{19}$~eV (red, large stars).
\label{skymap2}}
\end{figure}

In Fig.~\ref{skymap1}, we show a skymap in equatorial coordinates of
the arrival directions of the UHECR used in the analysis below.
An inspection by eye indicates an overdense region around and
south the AGASA triplet as well as several underdense regions or
voids. In Fig.~\ref{skymap2}, we show for comparison a skymap with
events from those two experiments (Hires and SUGAR) which published
also arrival directions below $E'=4\times 10^{19}$~eV. Again an
inspection by eye indicates that the addition of low-energy data appears 
to make the sky map more isotropic. In the next section, we perform a
statistical analysis to deduce the typical angular scales of excess
correlations visible for $E'=4\times 10^{19}$~eV.

\section{Autocorrelation analysis}
\label{analysis}

We use as our statistical estimator for possible deviations
from an isotropic distribution of arrival directions the angular two-point
auto-correlation function $w$. We define $w$ as function of the
angular scale $\delta$ as   
\be
 w(\delta) = \sum_{i=1}^N\sum_{j=1}^{i-1} \Theta(\delta-\delta_{ij}) \,,
\ee
where $\Theta$ is the step function, $N$ the number of CRs
considered and
$
\delta_{ij} = {\rm acos}\left( \cos\rho_i\cos\rho_j + 
              \sin\rho_i \sin\rho_j \cos(\phi_i - \phi_j) \right)
$
 is the angular distance between the two cosmic rays
$i$ and $j$ with coordinates $(\phi,\rho)$  on the sphere.
Having performed a large sample of Monte Carlo
simulations, we call the (formal) chance probability $P(\delta)$
to observe a larger value of the autocorrelation
function $w(\delta)$ the fraction of simulations with $w>w^\ast$, where
$w^\ast$ is the observed value. We would like to warn the reader at
this point that we have not fixed a priori our search and cut
criteria. Thus the obtained probabilities are only indicative. But
they can be used in particular to compare  for different data sets the
relative likelihood to observe the signal as chance fluctuation.

\begin{figure}
\epsfig{file=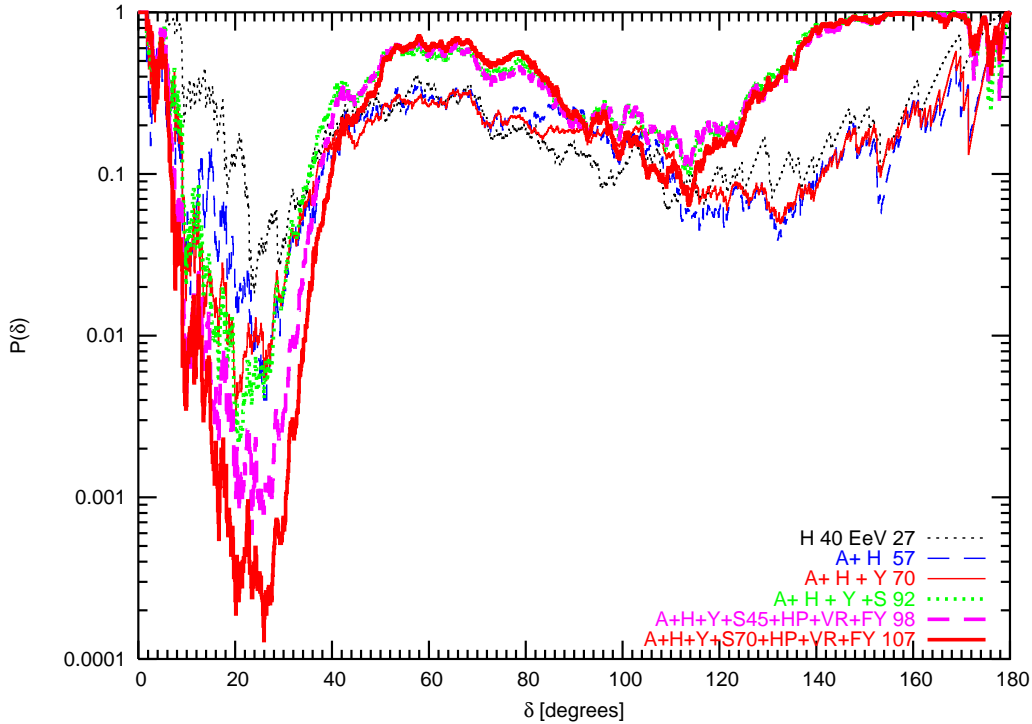,width=0.7\textwidth,angle=270}
\caption{
Chance probability $P(\delta)$ to observe a larger value of the autocorrelation
function as function of the angular scale  $\delta$ for different
combinations of experimental data; label of experiments as in Fig.~1. 
\label{p_ch_ex}}
\end{figure}

In Fig.~\ref{p_ch_ex}, we show the chance probability $P(\delta)$ as
function of the angular scale $\delta$ for different combinations of
experimental data. The chance probability $P(\delta)$ shows already a 
$2\sigma$ minimum around 20--30 degrees using only the 27~events of the HiRes
experiments with $E'\geq 4\times 10^{19}$~eV.
Adding more data, the signal around $\delta=25^\circ$
becomes stronger, increasing from $\sim 2\sigma$ for 27~events to $\sim
3.5\sigma$ for 107~events. It is comforting that the position of the
minimum of $P(\delta)$ is quite stable adding more data and every additional 
experimental dataset contributes to the signal. Moreover, 
autocorrelations at scales smaller than $25^\circ$ become more significant  
increasing the dataset. However,
we warn the reader again that we have not constrained ourselves a priori to
search for autocorrelations at $\delta=20^\circ$ and hence the numerical
values of the chance probability found should not be taken at face value.

To understand better how the search at arbitrary angular scales
influences the significance of our signal we have calculated 
the penalty factor\footnote{For a discussion of the use of penalty
factors see e.g Ref.~\cite{p}.} for the scan of $P(\delta)$ over
$\delta$. The penalty factor increases for increasing resolution
$\Delta\delta$ of the angular scale $\delta$, but reaches an asymptotic
value for $\Delta\delta\to 0$. The numerical value of the penalty factor
found by us in the limit $\Delta\delta\to 0$ varies
between 6 for the HiRes data set alone and 30 for the combination
of all data. Since the energy cut we use is determined by the one chosen 
in Ref.~\cite{cris}, no additional penalty factor for the energy has 
to be included. We conclude therefore that the true probability to
observe a larger autocorrelation signal by chance is $P\approx
3\times 10^{-3}$ for the complete data set.

\begin{figure}
\epsfig{file=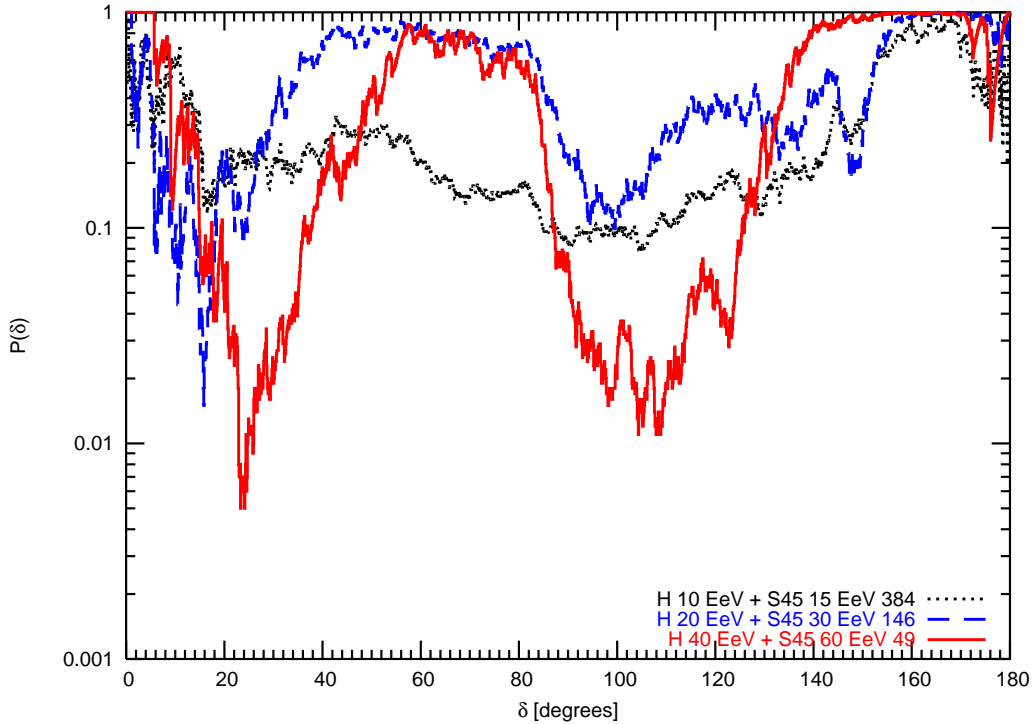,width=0.7\textwidth,angle=270}
\caption{
Chance probability $P(\delta)$ to observe a larger value of the autocorrelation
function as function of the angular scale  $\delta$ for different
cuts of the rescaled energy $E'$: black $E'\geq 1\times 10^{19}$~eV,
blue $E'\geq 2\times 10^{19}$~eV and red line $E'\geq 4\times
10^{19}$~eV.
\label{p_ch_E}}
\end{figure}

In Fig.~\ref{p_ch_E}, we show again the chance probability $P(\delta)$ as
function of the angular scale $\delta$, but now for different values of the
minimal event energy included in the analysis.  Again we can use only
Hires and SUGAR data, since for the other experiments no arrival
directions for events below $E'=4\times 10^{19}$~eV are
published. Moreover, the energy bin size we use is dictated by the
one chosen in Ref.~\cite{cris}. 
The addition of data with energy  below $E'\approx 4\times 10^{19}$~eV
reduces the significance of the minimum of $P(\delta)$. This could have
various reasons: First, the 
interaction length of the UHECR primaries can increase with decreasing
energy, as in the case of protons or nuclei. Then, the projection on
the two-dimensional skymap averages out more and more three-dimensional
structures. Second, deflections in the Galactic and extragalactic
magnetic fields destroy for lower energies more and more correlations. 
Note that the autocorrelation signal would not disappear lowering the
energy threshold if it would be caused solely by an incorrect
combination of the exposure of different experiments.
In Fig.~\ref{p_ch_E}, there is also a prominent minimum of $P(\delta)$ 
around $120^\circ$ visible. This angular scale corresponds to the 
distance between overdense spots and is always for other datasets in 
Fig.~\ref{p_ch_ex} visible, but much less significant.

\begin{figure}
\epsfig{file=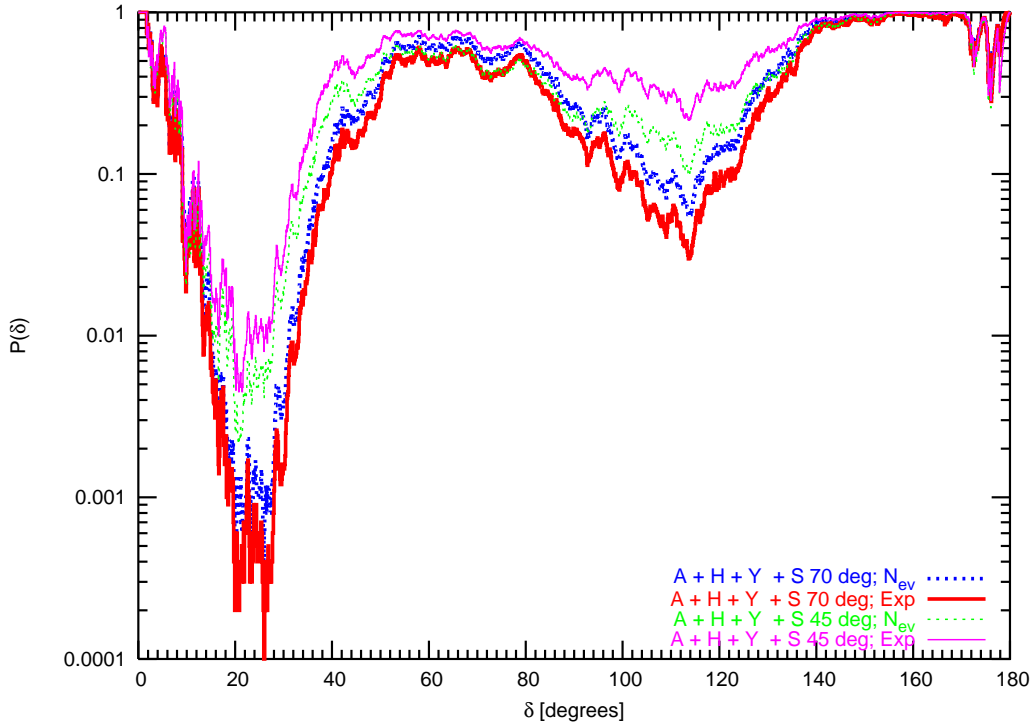,width=0.7\textwidth,angle=270}
\caption{
Chance probability $P(\delta)$ to observe a larger value of the 
autocorrelation
function as function of the angular scale  $\delta$ combining the AGASA, 
HiRes, Yakutsk and SUGAR experiments according to the number of observed 
events (dashed lines labeled $N_{ev}$) and according to the exposure of the 
individual experiments (solid lines labeled Exp).
Data of SUGAR zenith angles up to 45 degrees and up to 70 degrees are
shown with thin   
and thick solid lines, respectively.
\label{p_ch_exp}}
\end{figure}

In order to combine the various experiments in a single data set as shown in 
Figs.~\ref{p_ch_ex} and \ref{p_ch_E}, we have estimated the relative exposure 
of each experiment by the number of observed events above a chosen
energy threshold. We can check this procedure for those experiments
for which we can cut the data set on the same energy, i.e. for AGASA,
HiRes, Yakutsk and SUGAR. For this check, we fix the  total number of
events to the observed one, which is 92 events using  
zenith angle $\theta<45^\circ$ and 101 using $\theta<70^\circ$ for SUGAR.
Then we choose the event number of each experiment according to its exposure.
For instance, for SUGAR with zenith angle $\theta<45^\circ$ we use 32
HiRes, 27 AGASA, 12 Yakutsk, and 21 SUGAR events. Note that the
observed numbers of events are 27 HiRes,  30 AGASA,  13 Yakutsk, and
22 SUGAR event, i.e. the differences are compatible with pure
statistical (Poisson) fluctuations.   
The results of this comparison are presented in Fig.~\ref{p_ch_exp},
combining the AGASA, HiRes, Yakutsk with  SUGAR data with zenith
angles up to 45 degrees and up to 70 degrees, respectively. The chance
probability calculated with both recipes gives similar results. In one
case the exposure method gives slightly stronger and in the other case
slightly weaker results.

We have also checked how our result depends on the systematic energy
difference  between the AGASA and HiRes stereo experiments. Up to now,
we have assumed that the energies of the HiRes stereo data have no
systematic shift relative to the HiRes mono  
data~\cite{c2cr}. However, because the normalization of the HiRes
stereo spectrum is still under discussion, we have checked as extreme
possibility that the HiRes stereo 
energy scale is shifted upwards by 20\%. In this case, the  AGASA data
should be shifted down only by 10\%, while the Yakutsk and SUGAR data 
should be shifted by 26.5\%.  The total number of events in the sample
will increase  from 107 to 143.  The resulting  
probability  $P(\delta)$ as function of the angle $\delta$ is similar
to one shown in Fig.\ref{p_ch_ex}. Thus we conclude that our results
are not strongly sensitive to the exact value of the systematic energy
difference between the AGASA and the HiRes experiments. Note also that
once the relative energy scales of the experiments are  
fixed, our results do not depends on value of the ``true'' energy
scale, i.e. on the  
global shift of all experiments together up and down in energy.

\section{Discussion}

Our results, if confirmed by future independent data sets, have
several important consequences. 

Firstly, anisotropies on intermediate angular scales constrain the
chemical composition of UHECRs. Iron nuclei propagate in the 
Galactic magnetic field in a quasi-diffusive regime at $E=4\times
10^{19}$~eV and all correlations would be smeared out on scales as small as
observed by us. Therefore, models with a  dominating extragalactic
iron component at the highest energies are disfavored by 
anisotropies on intermediate angular scales. 

Secondly, the probability that small-scale clusters are indeed from
point sources will be reduced if the clusters are in regions with an
higher UHECR flux. For example, the AGASA triplet is located in an
over-dense spot (cf. map in Fig.~\ref{skymap1}) and the probability to
see a cluster in this region by chance is increased. In contrast, the 
observation of clusters in the "voids" of Fig.~\ref{skymap1} would be less
likely by chance than in the case of an UHECR flux without medium
scale anisotropies. 

However, the most important consequence of our findings is the prediction
that astronomy with UHECRs is possible at the highest energies. The
minimal energy required seems to be around $E'=4\times 10^{19}$~eV, 
because, as one can see from Figs.~\ref{skymap2} 
and \ref{p_ch_E}, at lower energies UHECR arrive more and more isotropically. 
This trend is expected, because at lower energies both deflections 
in magnetic fields and the average distance $l$ from which UHECRs can
arrive increase. 
Since the two-dimensional skymap corresponds to averaging all 
three-dimensional structures (with typical scale $L$) over the distance
$l$, no anisotropies are expected for $l\gg L$. Thus, if the signal
found in this analysis will be confirmed it has to be related to the local
large scale structure. 

Finally, we note that to check this signal an independent data set of order
$O(100)$ events with $E'>4\times 10^{19}$~eV is required. This agrees
roughly with the finding of Ref.~\cite{napoli} that around 100 events are 
needed to reject the hypothesis that the UHECR sources trace the galaxy 
distribution.

\section{Summary}
\label{sum}

We have found that the two-point autocorrelation function of UHECR
arrival directions has a broad maximum around 25 degrees. Combining
all publicly available data with energy $E'>4\times 10^{19}$~eV, the
formal chance probability that a stronger autocorrelation is obtained
from an isotropic distribution is around $P\sim 1\times 10^{-4}$. 
Both the signal itself and the exact value of the chance probability
have to be interpreted with care, since we have not fixed a priori
our search and cut criteria. Taking into account a penalty factor of
30 appropriate for our search at all angles $\delta\in[0:180^\circ]$,
we have estimated the true  probability as $P\approx 3\times 10^{-3}$. 
We have checked that the autocorrelation signal disappears lowering the
energy threshold, indicating that it is not caused solely by an incorrect
combination of the exposure of different experiments.
The autocorrelation signal found by us around 
$\delta=25^\circ$ should be tested with future, independent data sets
from HiRes, the Pierre Auger Observatory~\cite{PAO} and the Telescope
Array~\cite{TA}. If confirmed, it constrains the UHECR primary type 
together with the magnitude of extragalactic magnetic fields
and opens the door to astronomical studies with UHECRs.

\section*{Acknowledgments}
We are grateful to G\"unter Sigl, Igor Tkachev and Sergey Troitsky for helpful 
comments on the draft. We would like to thank Eric Armengaud and especially 
Pasquale Serpico for useful discussions.


\end{document}